\documentclass[twocolumn,showpacs,preprintnumbers,nofootinbib]{revtex4-1}
\usepackage{graphicx}  
\usepackage{dcolumn}
\usepackage{bm}
\usepackage{amsmath,amssymb,amsfonts}
\usepackage{latexsym}
\usepackage{color}

\usepackage[usestackEOL]{stackengine} 
\usepackage{multirow} 

\usepackage{hyperref}
\hypersetup{%
setpagesize=false,
 bookmarksnumbered=true,%
 bookmarksopen=true,%
 colorlinks=true,%
 linkcolor=blue,
 citecolor=red,
}

\def\dcp{\Delta \mathcal{A}_{\text{CP}}}

\def\bsg{\mathcal{B}(B\to X_s \gamma)}

\def\rtt{\rho_{tt}}
\def\irbb{\mbox{Im}(\rho_{bb})}
\def\rrbb{\mbox{Re}(\rho_{bb})}

\def\fbi{fb$^{-1}$~} 

\def\bzh{$bg \to b A \to b Z h$~}

\def\eeZH{$e^+ e^- \to Z^* \to A H$~}
\def\eehh{$e^+ e^- \to A H \to b \bar b hh$}

\def\nn    {\nonumber}

\begin{document}

\title{\boldmath
Electroweak baryogenesis via bottom transport:\\ complementarity between LHC and future lepton collider probes}

\author{Tanmoy Modak$^{1}$}
 \email{tanmoyy@thphys.uni-heidelberg.de}
 \author{Eibun Senaha$^{2,3}$}
 \email{eibun.senaha@tdtu.edu.vn}
\affiliation{
$^1$Institut f{\"u}r Theoretische Physik, Universit{\"a}t Heidelberg, 69120 Heidelberg, Germany\\
$^2$Theoretical Particle Physics and Cosmology Research Group, Advanced Institute of Materials 
Science, Ton Duc Thang University, Ho Chi Minh City, Vietnam\\
$^3$Faculty of Applied Sciences, Ton Duc Thang University, Ho Chi Minh City, Vietnam}

\bigskip

\begin{abstract}
We study the complementarity between the Large Hadron Collider (LHC) and future lepton colliders in 
probing electroweak baryogenesis induced by an additional bottom Yukawa coupling $\rho_{bb}$.
The context is general two Higgs doublet model (g2HDM) where such additional bottom Yukawa coupling 
can account for the observed baryon asymmetry of the Universe if $\mbox{Im}(\rho_{bb}) \gtrsim 0.058$.
We find that LHC would probe the nominal $\mbox{Im}(\rho_{bb})$ required for baryogenesis to some extent
via $bg \to bA \to bZh$ process if $300~\mbox{GeV}\lesssim m_A \lesssim 450$ GeV, where $A$ is the CP-odd scalar 
in g2HDM. We show that future electron positron collider such as International Linear Collider with $500$ 
GeV and 1 TeV collision energies may offer unique probe for the nominal $\mbox{Im}(\rho_{bb})$ via
$e^+ e^- \to Z^*\to A H$ process followed by $A,H \to b \bar b$ decays in four $b$-jets signature. 
For complementarity we also study the resonant diHiggs productions, which may give an
insight into strong first-order electroweak phase transition, via 
$e^+ e^- \to Z^*\to A H \to A h h$ process in six $b$-jets signature. We find that 
1 TeV collision energy with $\mathcal{O}(1)~\text{ab}^{-1}$ integrated luminosity
could offer an ideal environment for the discovery.
\end{abstract}

\maketitle

\section{Introduction}
The discovery of the 125 GeV Higgs boson ($h$)~\cite{Aad:2012tfa,*Chatrchyan:2012ufa} was a truly
watershed moment that established the standard model (SM) as a correct effective theory at electroweak scale.
While the SM has withstood all experimental tests so far,  cosmological problems such as 
baryon asymmetry of the Universe (BAU) and dark matter still remain open and a more fundamental theory must exist in nature.

For the BAU generation, one has to satisfy so-called Sakharov's 
conditions~\cite{Sakharov:1967dj}: (i) baryon number violation, (ii) C and CP violation,
and (iii) departure from thermal equilibrium.  One of the compelling ideas to explain
BAU is \textit{electroweak baryogenesis} (EWBG)~\cite{Kuzmin:1985mm} 
(for reviews, see 
Ref.~\cite{Rubakov:1996vz,*Funakubo:1996dw,*Riotto:1998bt,*Trodden:1998ym,*Quiros:1999jp,
*Bernreuther:2002uj,*Cline:2006ts,*Morrissey:2012db,*Konstandin:2013caa,*Senaha:2020mop}), whose core mechanism is already built-in even in the SM.
However, the observed parameters in the SM 
turn out to be inconsistent with successful EWBG due to the insufficient magnitude of 
CP violation and absence of a first-order electroweak phase transition (EWPT). 
Generally, various new-physics models are conceivable to circumvent those two issues. Among them, a general
two-Higgs-doublet model (g2HDM)~\cite{Djouadi:2005gi,*Branco:2011iw} is one of the most attractive models from the viewpoints of renormalizability, generality, and
testability. 
It is shown that extra Yukawa couplings of the second and third generation quarks and leptons, 
which can be complex and flavor violating, could provide CP violation sufficient for BAU. The 
most efficient EWBG scenario would be a case that the top quark has the $\mathcal{O}(1)$ 
extra Yukawa coupling, followed by a case in which the sizable top-charm-changing
Yukawa coupling is present~\cite{Fuyuto:2017ewj}. Thoroughgoing study of those collider signatures can be found in
Refs.~\cite{Kohda:2017fkn,Hou:2018zmg,Hou:2019qqi,Hou:2019gpn,Hou:2019mve,Ghosh:2019exx,Hou:2020tnc,Hou:2018uvr}. 

Under generous assumptions for bubble wall profiles, the bottom quark could also drive
the sufficient BAU if the size of the extra bottom Yukawa coupling is larger than the
SM bottom Yukawa coupling to some degree. This bottom-Yukawa-driven EWBG can be significant in a
case that the aforementioned Yukawa couplings in the up-type quark sector happen to be 
real or tiny. In Ref.~\cite{Modak:2018csw,*Modak:2020uyq}, 
the present authors studied phenomenological
consequences of the bottom-Yukawa-driven EWBG in detail assuming that
both extra top and bottom Yukawa couplings are present but the former is real
and the latter is roughly twice larger than a necessary bare minimum for BAU. It was 
found that Large Hadron Collider (LHC) with 1000 $\text{fb}^{-1}$ integrated luminosity could examine
the scenario, primarily via the process \bzh with final states
comprising of 3$b$-jets and a lepton pair, where $A$ is the CP-odd scalar 
and $h$ is the 125~GeV Higgs boson in the g2HDM.
In Ref.~\cite{Modak:2019nzl} it was also shown that the $bg\to bA \to b ZH$ process
would provide a sensitive test for the case of $m_A > m_H + m_Z$, where $H$ is the CP-even heavy scalar.
While these processes provide a unique probe to bottom-Yukawa-driven EWBG, they become insensitive
if $m_A < m_Z + m_h$ and/or $m_A < m_Z + m_H$.
Furthermore, if $m_A > 2 m_t$, an achievable significance
diminishes if the extra top Yukawa coupling is $\mathcal{O}(1)$.  

In this work, we further pursue the bottom-driven EWBG scenario with
particular emphasis on complementarity between LHC 
and the International Linear Collider (ILC). After taking the theoretical and
experimental constraints into accounts, we investigate a discovery 
potential of the EWBG scenario assuming a necessary bare minimum of 
the extra bottom Yukawa coupling and absence of the extra top Yukawa coupling,
which is diagonal parameter space investigated in Ref.~\cite{Modak:2020uyq}.
In this scenario, we examine the \bzh process at the LHC, and compare with the 
results in Ref.~\cite{Modak:2020uyq}. 
We also proceed to study detectability 
of EWBG signatures at the ILC assuming 500 GeV and 1 TeV center-of-mass (CM) energies.
We consider the process $e^+e^-\to AH$ with the $A/H\to b\bar b$ decay as well as the $H\to hh$ decay, 
leading to 4$b$-jets and 6$b$-jets final states, respectively.

The paper is organized as follows. In Sec.~\ref{formpar} we outline the model framework and 
the available parameter space for our study. Sec~\ref{bZh} is dedicated for finding
discovery prospect of the \bzh process. In Sec~\ref{eeZHprod} we discuss sensitivity of  
$e^+e^-\to AH\to 4b$. 
We also study the $e^+e^-\to AH\to Ahh$ process and the corresponding vertex correction for the trilinear $Hhh$ coupling.
We summarize our results with some discussions in Sec.~\ref{discu}.

\section{Formalism and Parameter space}\label{formpar}

\subsection{Formalism}\label{frame}
The most general CP-conserving g2HDM potential in the Higgs basis can be written as~\cite{Davidson:2005cw, Hou:2017hiw}
\begin{align}
 & V(\Phi,\Phi') = \mu_{11}^2|\Phi|^2 + \mu_{22}^2|\Phi'|^2
            - (\mu_{12}^2\Phi^\dagger\Phi' + \text{h.c.}) \nn \\
 & \quad + \frac{\eta_1}{2}|\Phi|^4 + \frac{\eta_2}{2}|\Phi'|^4
           + \eta_3|\Phi|^2|\Phi'|^2  + \eta_4 |\Phi^\dagger\Phi'|^2 \nn \\
 & + \left[\frac{\eta_5}{2}(\Phi^\dagger\Phi')^2
     + \left(\eta_6 |\Phi|^2 + \eta_7|\Phi'|^2\right) \Phi^\dagger\Phi' + \text{h.c.}\right],
\label{pot}
\end{align}
where the parameters $m_{11}^2$, $m_{12}^2$, $m_{22}^2$ and $\lambda_i$s are all real.
We consider Higgs basis where the vacuum expectation value $v(=246.22~\text{GeV})$ arises from the
doublet $\Phi$, \textit{i.e.}, $\left\langle \Phi\right\rangle =(0,v/\sqrt{2})^T$,
whereas $\left\langle \Phi'\right\rangle =(0,0)^T$,  assuming $\mu_{22}^2 > 0$.
The minimization condition with respect to the CP-even neutral component of $\Phi$ yeilds $\mu_{11}^2=-\frac{1}{2}\eta_1 v^2$, 
while that of $\Phi'$ gives $\mu_{12}^2=\frac{1}{2}\eta_6 v^2$. 
The mixing angle between $h$ and $H$ can be expressed as~\cite{Davidson:2005cw, Hou:2017hiw}
\begin{align}
 \cos^2\gamma = \frac{\eta_1 v^2 - m_h^2}{m_H^2-m_h^2},~\quad \quad \sin{2\gamma} = \frac{2\eta_6 v^2}{m_H^2-m_h^2}.
\end{align}
In the following we use shorthand $c_\gamma$ and $s_\gamma$
for $\cos\gamma$ and $\sin\gamma$ respectively while in the alignment limit
$c_\gamma \to 0$ and $s_\gamma \to -1$.

The Yukawa sector of the g2HDM is given by~\cite{Davidson:2005cw}
\begin{align}
\mathcal{L} = 
&-\frac{1}{\sqrt{2}} \sum_{F = U, D, L'}
 \bar F_{i} \bigg[\big(-\lambda^F_{ij} s_\gamma + \rho^F_{ij} c_\gamma\big) h \nn\\
 &+\big(\lambda^F_{ij} c_\gamma + \rho^F_{ij} s_\gamma\big)H -i ~{\rm sgn}(Q_F) \rho^F_{ij} A\bigg]  P_R\; F_{j}\nn\\
 &-\bar{U}_i\left[(V\rho^D)_{ij} P_R-(\rho^{U\dagger}V)_{ij} P_L\right]D_j H^+ \nn\\
 &- \bar{\nu}_i\rho^L_{ij} P_R \; L_j H^+ +{\rm H.c.},\label{eff}
\end{align}
where $P_{L,R}\equiv (1\mp\gamma_5)/2$, $V$ is CKM matrix, $i,j = 1, 2, 3$ are generation indices,
and $U=(u,c,t)^T$, $D = (d,s,b)^T$, $L=(e,\mu,\tau)^T$ and $\nu=(\nu_e,\nu_\mu,\nu_\tau)^T$ are column vectors in the flavor space.
The matrices $\lambda^F_{ij}\; (=\sqrt{2}m_i^F\delta_{ij}/v)$ are real and diagonal,
while $\rho^F_{ij}$ are in general complex and non-diagonal. 
It is pointed out in Ref.~\cite{Fuyuto:2019svr} that electric dipole moment (EDM) of the
electron could be suppressed if the diagonal elements of $\rho_{ij}^F$ follow the similar
hierarchal structures of the SM Yukawa couplings, i.e., $|\rho_{ee}/\rho_{tt}|\sim \lambda_e/\lambda_t$, 
which tempts us to conjecture $|\rho_{ii}/\rho_{jj}|\sim \lambda_i/\lambda_j$ for all the flavor indices.
We however consider somewhat offset parameter space motivated by the successful $\rho_{bb}$-EWBG mechanism
in which $\irbb = 0.058(\gtrsim\lambda_b\simeq 0.024)$. Circumvention of the electron EDM constraint
in this scenario will be addressed in Sec.~\ref{const}.

Here we should note that $h$, $H$, and $A$ are not CP eigenstates any more when including loop 
corrections that break CP through $\text{Im}(\rho_{ij})$. However, the loop corrections are small enough to 
regard the neutral Higgs bosons as the CP as well as mass eigenstates.

For all practical purposes we turn off all $\rho_{ij}$ except for $\rho_{bb}$, however their impact will be discussed in Sec.~\ref{discu}. 

Primary motivation of this article is to probe the nominal value $\irbb = 0.058$~\cite{Modak:2018csw} required for $\rho_{bb}$-EWBG. 
In general, LHC would offer exquisite probe via \bzh process if $\irbb \gtrsim 0.15$~\cite{Modak:2020uyq} but the process requires
$m_A > m_Z + m_h$. The process $bg\to bA \to b ZH$ would also offer sensitive probe if $m_A > m_H + m_Z$~\cite{Modak:2019nzl}. 
We note that the dependence of the $AZ h$ and $A Z H$ couplings on the mixing angle $\gamma$ can be found from~\cite{Djouadi:2005gi,*Branco:2011iw}
\begin{align}
 \frac{g_2 }{2 c_W}Z_\mu \left[c_\gamma (h \partial^\mu A  - A \partial^\mu h) 
 -s_\gamma(H \partial^\mu A - A \partial^\mu H)\right],
\label{zhlagra}
\end{align}
where $c_W$  and $g_2$ are the Weinberg angle and the $SU(2)_L$ gauge coupling respectively.
As discussed in Ref.~\cite{Modak:2018csw,Modak:2020uyq}, the nonzero $\gamma$ could 
have non-negligible impacts on $\rho_{bb}$-EWBG. From the interactions (\ref{eff}) and 
(\ref{zhlagra}), one can see that the production $bg\to b A$ does not depends on $\gamma$ and 
the decays $A\to Z H$ and $A \to Zh$ are scaled by $s_\gamma$ and $c_\gamma$, respectively.
In the vicinity of the alignment limit $\gamma=-\pi/2$, the \bzh process would provide more 
sensitive probe of the mixing angle through $c_\gamma$. While \bzh process can exclude the nominal $|\irbb| = 0.058$ at HL-LHC
if $m_A\sim 300$ GeV, it fails to probe the nominal $\irbb$ above $m_A > 2 m_t$ if $\rtt\sim 0.5$~\cite{Modak:2020uyq}.
Here we shall revisit potential of \bzh process to probe nominal $\irbb$ for scenarios where $m_A > 2 m_t$ but for vanishingly small $\rtt$.

The \bzh process would become insensitive for $m_A < m_h + m_Z$. In such scenarios future lepton colliders
such as ILC or FCCee would offer unique probe for $\rho_{bb}$-EWBG  via $e^+e^- \to Z^* \to A H$ process~\cite{Baer:2013cma,*AguilarSaavedra:2001rg,*Aoki:2009ha}
followed by $A/H\to b \bar b$ decays i.e., in four $b$-jets signature.
The signature would also receive contribution from $\rho_{bb}$ induced $e^+ e^- \to Z^*\to b\bar b A/H$ 
process if $A,H$ decays to $b \bar b$.
We remark that a similar search $pp\to Z^*\to A H \to b \bar b b \bar b$ at the 
LHC would suffer from overwhelming QCD multi-jets backgrounds, which prevents us from probing our scenario. 

Given the fact that the strong first-order EWPT needs $\mathcal{O}(1)$ Higgs quartic couplings, 
triple Higgs couplings $\phi_i\phi_j\phi_k$ could be potentially large.  A sensitive probe for $Hhh$
coupling is possible via $e^+ e^- \to Z^*\to A H \to A h h$ process (see Ref.~\cite{Djouadi:1999gv} for similar discussion). We study this process
in six $b$-jets signature. The final state signature would receive contribution from
$e^+e^- \to b \bar b H \to b \bar b hh$\footnote{Similar final signature has been discussed 
in the context of a softly $Z_2$-broken 2HDM in Ref.~\cite{Ahmed:2021kgl}} if both the $h$ decays to $b \bar b$. 
The $Hhh$ coupling is defined as the coefficient of the $h^2H$ term in the Higgs potential, from which it follows that~\cite{Hou:2019qqi}
\begin{align}
\lambda_{Hhh} & =  \frac{v}{2} \bigg[3 c_\gamma s_\gamma^2 \eta_1+ c_\gamma (3c_\gamma^2 - 2) \eta_{345}\nn\\
 & \quad\quad + 3 s_\gamma( 1-3c_\gamma^2)\eta_6+3s_\gamma c_\gamma^2 \eta_7\bigg]\label{lHhh},
\end{align}
 with $\eta_{345}=\eta_3+\eta_4+\eta_5$.
For small $c_\gamma$, $\lambda_{Hhh}$ is reduced to
\begin{align}
 \lambda_{Hhh}
 &\simeq-\frac{c_\gamma}{2v}\bigg[m_H^2-4\mu_{22}^2 +3c_\gamma \eta_7+\mathcal{O}(c_\gamma^2)\bigg],
 \label{lHhhapp}
\end{align}
which implies that $\lambda_{Hhh} \to 0$ as $c_\gamma\to 0$.
The approximate expression (\ref{lHhhapp}) does not differ from the exact one (\ref{lHhh}) 
by more than about 1.5\% in our benchmark points (BPs) described below. We also notice
that $\lambda_{Hhh}$ is always negative in our chosen BPs, which could be important when discussing one-loop corrections.
We primarily focus on tree-level $Hhh$ coupling
however we will discuss higher-order corrections
to $\lambda_{Hhh}$ and its impact on strong first-order EWPT in Sec.~\ref{eeHtohh}.
A probe for $Hhh$ coupling in the context of $\rho_{bb}$-EWBG 
would be indeed possible at the LHC via $b \bar b \to H \to h h$ and $bg\to bH \to b h h$. However 
we have checked that such processes are beyond the scope of the HL-LHC for nominal value $|\irbb| =0.058$
primarily due to overwhelming SM QCD background such as multi-jets and $t\bar t$+jets.

\subsection{Constraints and parameter space}\label{const}
Let us find the allowed parameter space for $m_A$, $m_H$ and $m_{H^\pm}$ such that EWBG is possible.
As widely known, $\eta_i v^2$, where $\eta_i$ are some linear combinations of $\eta$'s whose
magnitude is $\mathcal{O}(1)$, should be greater than $\mu_{22}^2$ in order to induce the 
strong first-order EWPT, leading to lower bounds of the heavy Higgs bosons.
On the other hand, since the quartic couplings are enforced to satisfy perturbativity
and tree-level unitarity, their sizes cannot exceed certain values, 
e.g., $4\pi$, which sets upper bounds of the heavy Higgs bosons.\footnote{
We also evaluate a scale at which one of $\eta$'s exceeds $4\pi$, where the theory starts to enter non-perturbative regime. Using one-loop renormalization group equations (RGEs) with $m_A$ as an initial value, it is found that $\Lambda_{\text{non-perturb}}=(2.0, 1.7, 2.5)$ TeV in the 3 benchmark points shown in Table \ref{bench}, respectively. Those scales could be roughly doubled if two-loop RGEs are used (see, e.g., Ref.~\cite{Dorsch:2016nrg}).
}
Therefore,
typical mass window for the strong first-order EWPT would be $m_{A,H, H^\pm}\in[200, 600]$ GeV.

The parameters in Eq.~\eqref{pot} are required to satisfy perturbativity, tree-level
unitarity and vacuum stability conditions, for which we utilized the public tool \texttt{2HDMC}~\cite{Eriksson:2009ws}.
We choose three BPs summarized in Table~\ref{bench} that satisfy 
aforementioned three theoretical constraints, electroweak precision measurements, and strong first-order EWPT as needed for EWBG.

\begin{table*}[htbp]
\centering
\begin{tabular}{c |c| c| c| c | c | c| c | c |c| c |c | c | c| cc}
\hline
BP & $\eta_1$ &  $\eta_2$   &  $\eta_3$   & $\eta_4$  & $\eta_5$ & $\eta_6$  & $\eta_7$ 
& \Centerstack{ $m_{H^\pm}$}  & \Centerstack{ $m_A$} & \Centerstack{ $m_H$} & $\mu_{22}^2/v^2$& $c_\gamma$ & $s_\gamma$ &  $\lambda_{Hhh}$ (GeV)\\
\hline
$a$        & 0.263  & 3.768 & 3.829 & $-2.27$  & 0.022  & $-0.054$ & 0.404  & 341  & 216  & 220  & 0.000373 & 0.1  & $-0.995$  & $-10.95$\\
$b$        & 0.271  & 3.265 & 5.968 & $-3.005$ & 0.0135 & $-0.132$ & 2.291  & 431  & 307  & 310  & 0.078    & 0.1  & $-0.995$  & $ - 23.62$\\
$c$        & 0.297  & 3.3   & 4.589 & 1.409  & 0.734  & $-0.395$ & 2.753  & 435  & 457  & 506  & 0.818    & 0.1  & $-0.995$  & $ -21.71$ \\
\hline
\hline
\end{tabular}
\caption{Parameter values of the three benchmark points. The masses $m_{H^\pm}$, $m_A$ and  $m_H$ are given in GeV.}
\label{bench}
\end{table*}

Having fixed the BPs, we now turn our attention to constraints on $\irbb$. 
There exist several indirect and direct searches that can constrain the parameter
space for $\irbb$. For nonvanishing $\rho_{tt}$, $\irbb$ receives meaningful
constraints from the branching ratio measurement of $B \to X_s \gamma$ ($\bsg$) and
the asymmetry of the CP asymmetry between the charged and neutral $B\to X_s \gamma$ decays ($\dcp$)~\cite{Modak:2018csw, Modak:2020uyq}.
However as we focus on parameter space where $\rho_{tt}$ is small, such constraints practically allow an order of magnitude larger
$\irbb$ than that of the nominal value required for $\rho_{bb}$-EWBG. Therefore we do not discuss such constraint here and redirect
readers to Refs.~\cite{Modak:2018csw,Modak:2020uyq} for further details.

\begin{figure}[t]
\centering
\includegraphics[width=.4 \textwidth]{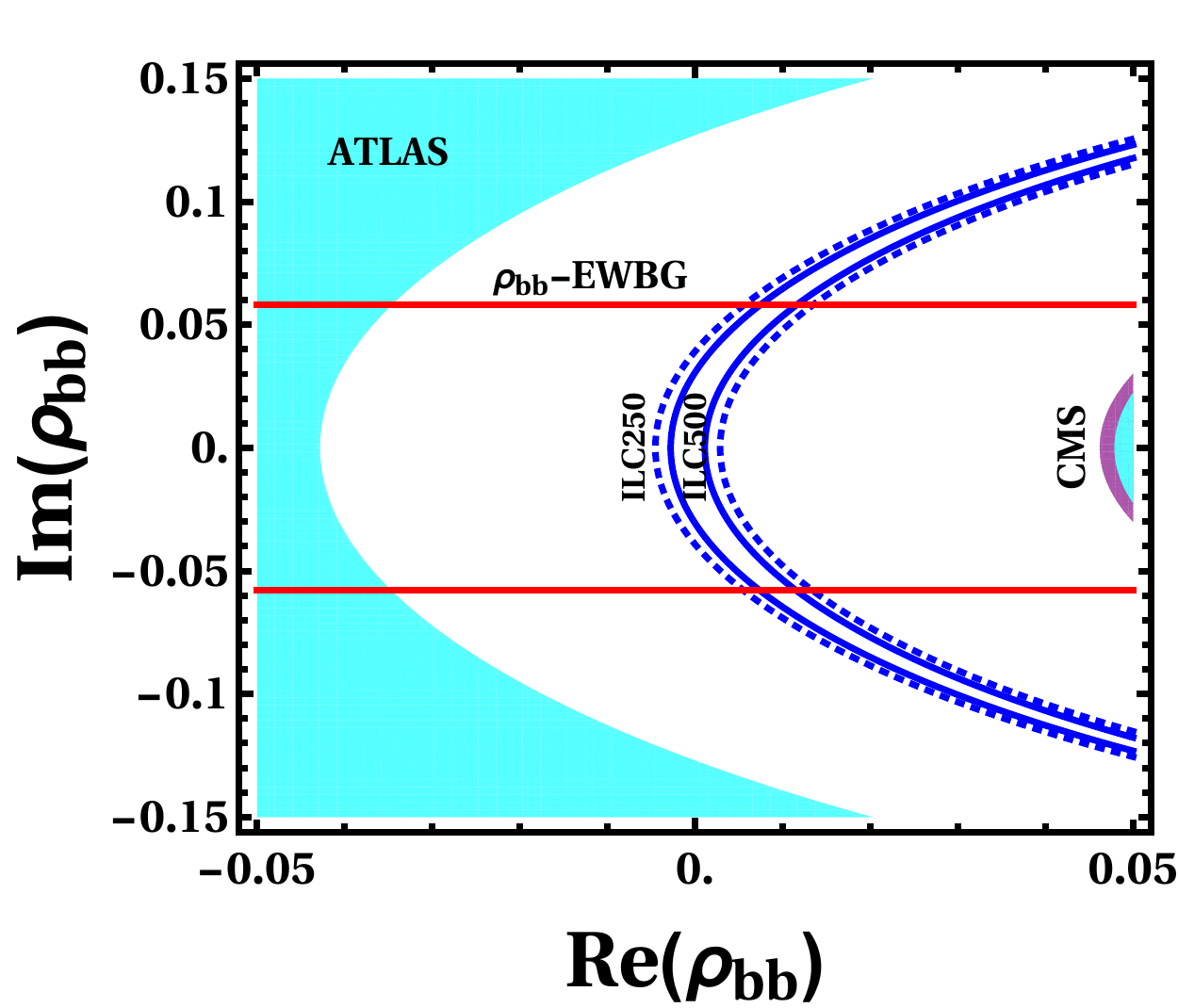}
\caption{Allowed parameter space for $\rho_{bb}$ from Higgs coupling measurements along
with future projections of ILC 250 GeV and 500 GeV run in the case of $c_\gamma=0.1$. $|\text{Im}(\rho_{bb})|>0.058$ corresponds to the region of $\rho_{bb}$-EWBG.}
\label{paramspace}
\end{figure}

The Higgs signal strength measurements by ATLAS and CMS would however provide some constraints primarily due to our 
choice of $c_\gamma =0.1$, which should be clear from the Yukawa couplings of $h$ in Eq.~\eqref{eff}. Although no combined
analysis has been performed, both CMS and ATLAS collaborations have updated the $h$ boson
coupling measurements with full Run 2 data~\cite{CMS:2020gsy,ATLAS:2020qdt}. 
The central value and $1\sigma$ error bar is provided for the coupling 
modifier $\kappa_b$, which is defined as the ratio between the observed and
SM partial rates (see Refs.~\cite{CMS:2020gsy,ATLAS:2020qdt} for its definition), 
by both the collaboration. 
We identify $\kappa_b$ from Eq.~\eqref{eff} as:
\begin{align}
 |\kappa_b| = \sqrt{\left(-s_\gamma+\frac{c_\gamma \rrbb}{\lambda_b}\right)^2+\left(\frac{\irbb c_\gamma}{\lambda_b}\right)^2}\label{kapb},
\end{align}
where $\lambda_b =\sqrt{2} m_b/v$ with $m_b$ is  $\overline{\rm MS}$ mass of $b$ quark evaluated at $m_h$.
The CMS found $\kappa_b=1.18^{+0.19}_{-0.27}$~\cite{CMS:2020gsy} whereas 
ATLAS found $\kappa_b=0.98^{+0.14}_{-0.13}$~\cite{ATLAS:2020qdt}. Allowing $2\sigma$
error bars on these measurements we show these limits in Fig.~\ref{paramspace} 
in the $\rrbb$--$\irbb$ plane by purple (CMS) and cyan (ATLAS) shaded regions.
While finding the limits we simply symmetrized the error bars of CMS and ATLAS measurements.
For comparison we also overlay the nominal parameter space for $\rho_{bb}$--EWBG ($|\text{Im}(\rho_{bb})|>0.058$) by the red solid lines in Fig.~\ref{paramspace}.
It is clear that $\kappa_b$ measurements are not able to cover the entire $\rho_{bb}$-EWBG region. 
This is primary due to the fact that CP-violating term $\irbb$ does not interfere with the SM part, 
thereby being more suppressed by the mixing angle $c_\gamma$, as can be seen from Eq.~\eqref{kapb}.
It would be useful to compare the sensitivity of future $e^+e^-$ collider in probing $\kappa_b$.
In this regard we focus on the ILC, which is expected to measure $\kappa_b$ within $1.1\%$ and $0.58\%$~\cite{Fujii:2017vwa}
uncertainties at $1\sigma$ in its $\sqrt{s}=250$ GeV (denoted as ILC250) and, combined 250 GeV and 500 GeV data (denoted as ILC500). 
Allowing $2\sigma$ error we illustrate these limits in Fig.~\ref{paramspace} by blue dotted and solid lines respectively, where in both cases 
the white crescent shaped regions within the lines are allowed.
For comparison the HL-LHC is expected to measure $\kappa_b$ with 
$\approx 6\%$ accuracy~\cite{Cepeda:2019klc}, which we do not show in 
Fig.~\ref{paramspace}. It is clear that sufficient parameter space for $\rho_{bb}$--EWBG would survive
even after various precise measurements of $hbb$ coupling.

There also exist some heavy Higgs searches from ATLAS and CMS that also constrain $\irbb$.  
E.g., it was found~\cite{Modak:2019nzl,Modak:2020uyq} that the most relevant constraints arise from
heavy neutral Higgs boson production with at least one $b$-jet followed by $b\bar b$ decay~\cite{Sirunyan:2018taj} 
and, heavy charged Higgs searches $pp\to t(b)H^\pm$ with  $H^+/H^- \to t \bar b/ \bar t b$ decays~\cite{Aad:2021xzu,Sirunyan:2020hwv}~
(see also e.g. Refs.~\cite{Plehn:2002vy,Boos:2003yi}). 
As we primarily focus on parameter space where $|\irbb|\approx 0.058$ and, the fact that such searches
excludes $\irbb\gtrsim 0.25$~\cite{Modak:2019nzl,Modak:2020uyq} for the sub-TeV mass range, 
we refrain a detailed discussion of these here and redirect readers to Refs.~\cite{Modak:2019nzl,Modak:2020uyq}
for further discussion.

 Now we discuss EDM constraint on $\text{Im}(\rho_{bb})$ in light of the latest result of ACME Collaboration~\cite{Andreev:2018ayy}.
This constraint is so overwhelming that one cannot dodge it without relying on some mechanism in any EWBG scenarios in g2HDM.
As briefly mentioned below Eq.~(\ref{eff}), the electron EDM could be sufficiently
suppressed by the build-in cancellation mechanism. For that end, $\rho_{tt}$ and $\rho_{ee}$ have to be complex and echo the SM-like Yukawa hierarchy.
In the $\rho_{bb}$-EWBG scenario, however, $\rho_{tt}$ is real or small by 
assumption and the above solution space is the no-go zone. Nonetheless, it is
still possible to render the electron EDM small enough to avoid the ACME constraint
in concert with $\rho_{bb}$ and $\rho_{ee}$ though the cancellation does not manifest
any structure. We do not repeat the analysis here and refer the readers to Ref.~\cite{Modak:2020uyq} for more details.

\begin{table}[h]
\centering
\begin{tabular}{|c| c| c| c| c | c | c|}
\hline             
BP  & $b\bar b$ & $Zh$ & $ZH$ & $\Gamma_A$ (GeV) \\
\hline
\hline
$a$      &  1.00         & --                   & --   & 0.043\\
$b$      & 0.648      & 0.352                & --   & 0.095\\
$c$      & 0.301      & 0.699                & --   & 0.304 \\
\hline
\hline
\end{tabular}
\caption{The branching ratios and total widths of $A$ for the benchmark points. Here we assumed
$|\irbb|=0.058$ and set all $\rho_{ij}=0$. See text for details.}
\label{branch:A}
\end{table}

\begin{table}[h]
\centering
\begin{tabular}{|c|c|c|c|c|c|c|}
\hline
BP  & $b\bar b$ & $h h$ & $WW$ & $ZZ$ & $t\bar t$ & $\Gamma_H$ (GeV) \\
\hline
\hline  
$a$  & 0.658 & --    & 0.244 & 0.098 & --    & 0.067\\
$b$  & 0.310 & 0.213 & 0.330  & 0.147 & --    & 0.199\\ 
$c$  & 0.128 & 0.041 & 0.462 & 0.222 & 0.147 & 0.790 \\  
\hline
\hline
\end{tabular}
\caption{The branching ratios and total widths of $H$ for the benchmark 
points for $|\irbb|=0.058$ with all other $\rho_{ij}=0$.}
\label{branch:H}
\end{table}

Without significant improvements in experimental uncertainties all in all we remark 
that the nominal value  $|\irbb|= 0.058$ for $\rho_{bb}$-EWBG is likely
to survive all current and future measurements discussed in this section. For illustration
we take $|\irbb|= 0.058$ for all three BPs in our analysis. For all practical purposes we
set all $\rho_{ij}=0$ except for $\irbb$ however we shall return to the impact of turning
other $\rho_{ij}$ couplings in Sec.~\ref{discu}. Under the aforementioned assumption and 
neglecting tiny loop induced decays such as $A\to \gamma\gamma$, the CP-odd boson 
$A$ decays practically 100\% to $b\bar b$ for BP$a$, while additional decay mode $Zh$ 
are open and constitute about 35\% and 70\% for BP$b$ and BP$c$, respectively. 
The respective branching ratios for the three BPs are summarized in Table~\ref{branch:A}. 
For the CP-even heavy Higgs boson $H$, it primarily decays to $b\bar b$, followed by $WW$ 
and $ZZ$ in BP$a$. In BP$b$, the $b\bar{b}$ and $WW$ modes comprise about 30\% branching 
ratios, followed by $hh$ and $ZZ$. In BP$c$, $WW$ is the dominant decay mode, followed by $ZZ$. 
The $t\bar{t}$ channel is also kinematically accessible, which predominates over the $b\bar{b}$ and $hh$ modes.
In addition to the above decay modes, the decays such as $H\to \tau \tau$, $H\to c \bar c$, etc. 
would be turned on via nonzero $c_\gamma$, as can be seen from Eq.~\eqref{eff}.
Besides tiny loop-induced decays, the respective branching ratios of $H$ for the three BPs 
are given in Table~\ref{branch:H}. Here in both tables we consider branching ratios with 
three significant digits.

\section{The \bzh process}\label{bZh}
We first analyze the prospect of discovering nominal $\irbb$ required for EWBG via \bzh process at HL-LHC.
The process can be searched at the LHC via $pp\to b A +X \to b Zh +X$~\cite{Aaboud:2017cxo,*Sirunyan:2019xls,*Ferreira:2017bnx,*Coyle:2018ydo}
followed by $Z\to \ell^+ \ell^-$ ($\ell = e,~\mu$) and $h\to b \bar b$ i.e., in signature comprising
of a pair of same flavor opposite sign leptons (denoted as the $bZh$ process) and three $b$-tagged jets. The process requires that 
$m_A > m_Z + m_h$. Therefore BP$a$ for which $m_A < m_Z + m_h$ is out of the reach of LHC and we only focus on BP$b$ and BP$c$.
There exist several SM backgrounds such as $t\bar t+$jets, Drell-Yan+jets (DY+jets), $Wt+$jets, $t\bar tZ$+jets, $t\bar t h$, $tZ$+jets,
whereas subdominant contributions arise from four-top ($4t$), $t\bar t W$, $tWh$, $tWZ$ and $WZ$+jets. 
Backgrounds from $WW$+jets is negligibly small and hence not included. We remark that a search 
can also be performed via $h\to \tau \tau$ and $h\to \gamma \gamma$ modes, however, they are not as promising as $h\to b\bar{b}$.

We generate the signal and SM background event samples at leading order (LO) in $pp$ collision with $\sqrt{s}=14$ TeV CM energy 
by MadGraph5\_aMC@NLO~\cite{Alwall:2014hca} (denoted as MadGraph5\_aMC) with default NN23LO1 PDF 
set~\cite{Ball:2013hta} then interface with Pythia 6.4~\cite{Sjostrand:2006za} for 
hadronization and showering and, finally fed  into Delphes~3.4.2~\cite{deFavereau:2013fsa} for the fast detector simulation 
incorporating the default ATLAS-based detector card.
We follow MLM scheme~\cite{Mangano:2006rw,Alwall:2007fs} for the matrix element (ME) and parton shower merging.  
Note that we do not included  backgrounds from  the fake and non-prompt sources in our analysis which
are typically determined from data and are not properly modeled in the Monte Carlo simulations.
The effective model is implemented in FeynRules 2.0~\cite{Alloul:2013bka} framework.

The DY+jets background cross section
is adjusted to the NNLO QCD+NLO EW one by a factor 1.27, which is estimated by FEWZ 3.1~\cite{Li:2012wna,Hou:2017ozb}, while
the $t\bar t+$jets background is corrected up to NLO by the $K$ factor $1.36$~\cite{Alwall:2014hca}.
We also normalize the LO $t\bar t Z$,  $\bar tZ +$ jets, $t\bar t h$, $4t$ and  $t\bar{t} W^-$ ($t\bar{t} W^+$) cross sections 
to NLO ones by the $K$-factors 1.56~\cite{Campbell:2013yla}, 1.44~\cite{Alwall:2014hca}, 1.27~\cite{twikittbarh}, 
2.04~\cite{Alwall:2014hca} and 1.35 (1.27)~\cite{Campbell:2012dh} respectively,
while both $tWZ$ and $tWh$ are kept at LO. The LO $W^-Z+$jets background is normalized to NNLO by a factor 
2.07~\cite{Grazzini:2016swo}. We assume the same QCD correction factors for the charge conjugate processes $tZj$ and $W^+Z+$jets.
The signal cross sections are kept at LO.

\begin{table*}[hbt!]
\centering
\begin{tabular}{|c |c| c| c| c | c| c| c |c |c|}
\hline
 BP &  Signal (fb) & $t\bar t+$ jets  & $DY+$ jets   & $Wt+$ jets   &  $t\bar t Z$   & Others & Total Bkg.  (fb) \\
\hline
\hline
       $b$     &  0.064      & 0.270  & 0.702  & 0.404  & 0.017 & 0.01   & 1.403 \\
       $c$     &  0.025      & 0.108  & 0.139  & 0.024  & 0.007 & 0.002 &  0.281 \\ 

\hline
\hline
\end{tabular}
\caption{The signal and background cross sections (in fb) of the $bZh$ process 
after selection cuts for the respective BPs at $\sqrt{s}=14$ TeV LHC.
We have assumed $|\irbb|=0.058$ and set all other $\rho_{ij}=0$ for the signal process.
The subdominant backgrounds $4t$, $t\bar t W$, $tWh$, $tWZ$ and $WZ$+jets are added together and denoted as 
``Others''. The total background yield (Total Bkg.) is given in the last column.}
\label{sigbkgbzh}
\end{table*}

In order to find the prospect, we look for event topologies with
same flavor opposite sign lepton pair and at least three $b$-tagged jets.
To reduce backgrounds we apply following event selection cuts. 
The transverse momenta ($p_T$) of the leading and subleading leptons are required to be
$> 28$ GeV and $>25$ GeV respectively, while $p_T> 20$ for all the three $b$-jets. 
The pseudo-rapidity ($|\eta|$) for all the leptons and $b$-jets are needed to  satisfy $|\eta|<2.5$.
Moreover, the separation $\Delta R$ between the two leptons, any two $b$-jets and, a $b$-jet and a lepton should be $ \Delta R>0.4$.
The jets are reconstructed with anti-$k_T$ algorithm via default ATLAS-based detector card of Delphes~3.4.2.
We veto events with missing transverse energy ($E_T^{\rm{miss}}$) $> 35$ GeV to reduce the $t\bar t$+jets background.
We further require that the invariant mass of the same flavor opposite charge lepton pair ($m_{\ell\ell}$) should remain
within $ 76 < m_{\ell\ell} < 100$ GeV, i.e., in the $Z$ boson mass window. The invariant mass of two 
$b$-jets $m_{bb}$ in an event to remain within $|m_h-m_{bb}|< 25$ GeV.
As each event contains at least three $b$-jets more than one $m_{bb}$ combinations are possible;
the one closest to $m_h$ is selected to pass the $|m_h-m_{bb}|< 25$ GeV cut.
We finally require the invariant mass $m_{\ell\ell bb}$ constructed from the same flavor opposite charge lepton pair that 
pass the $76 < m_{\ell\ell} < 100$ GeV window and $b$-jets combination that passes the $|m_h-m_{bb}|< 25$ GeV selection 
to remain within $|m_A- m_{\ell\ell bb}|< 80$ GeV. 
Here we adopt the $b$-tagging and $c$- and light-jets rejection efficiencies ATLAS based detector card of Delphes~3.4.2.
The signal and background cross sections after the selection cuts for the BP$b$ and BP$c$ are summarized in Table.~\ref{sigbkgbzh}.

We now focus on the achievable significance at HL-LHC using the likelihood for a simple counting experiment~\cite{Cowan:2010js}
\begin{align}
  Z(n|n_\text{pr})= \sqrt{-2\ln\frac{L(n|n_\text{pr})}{L(n|n)}};\; L(n|\bar{n}) = \frac{e^{-\bar{n}} \bar{n}^n}{n !} \, \label{poisso},
\end{align}
where $n$ and $n_\text{pr}$ are observed and predicted events. 
For discovery, the signal plus background ($s+b$) is compared with the
background prediction ($b$) with the requirement $Z(s+b|b) > 5$, while for the exclusion we
demand $Z(b|s+b) > 2$~\cite{Cowan:2010js}. An evidence would require $Z(s+b|b) > 3$.
Utilizing the signal and background cross sections in Table~\ref{sigbkgbzh} 
we find that the achievable significance is $\sim 2.9\sigma$ for BP$b$ while $\sim 2.5\sigma$ for BP$c$ with 3000 fb$^{-1}$ integrated luminosity. 
Therefore we conclude that the discovery is beyond the 
scope of HL-LHC if the $\irbb$ close to its nominal value 0.058 required for $\rho_{bb}$-EWBG.
We note that $\irbb\sim 0.15$--0.2 is still allowed by current data for the sub-TeV mass
range as we have already discussed in previous section. We also remark that in BP$c$, for which
$m_A > 2 m_t$, discovery is well within the HL-LHC if one considers $\irbb\sim 0.15$. This is 
different from a scenario discussed in Ref.~\cite{Modak:2020uyq} (also referred to as BP$c$ there) 
in which $\mathcal{B}(A\to Zh)$ is suppressed due to the dominance of $\mathcal{B}(A\to t \bar t)$ 
induced by $|\rho_{tt}|=0.5$, hindering the significance from reaching the discovery level.

\section{The \eeZH production}\label{eeZHprod}
\subsection{The four $b$-jets signature}
In this section we investigate the potential for $eeAH$ process i.e., $e^+ e^- \to Z^*\to A H$ production
with $H/A \to b \bar b$ decays in four $b$-jets signature for two different $e^+e^-$ collision energy $\sqrt{s}= 500$ GeV and 1 TeV. 
The signature would also receive contribution from 
$e^+ e^- \to Z^* \to b \bar b A/H$ process for $A/H\to b \bar b$ decays, which we have included in our analysis. 
It is clear from Table~\ref{bench} that BP$a$ would be covered by $\sqrt{s}= 500$ GeV while BP$b$ and BP$c$ 
would require  $\sqrt{s}= 1$ TeV. 

Although the environment is clean, there indeed exist some SM backgrounds for this process.
The dominant backgrounds come from $t\bar t$, four-jets ($4j$) which includes $Zh$ productions, with 
subdominant contribution would arise from $ZZ$ background.
The events are generated as in previous section by MadGraph5\_aMC followed by showering
and hadronization in PYTHIA~6.4, and fed into Delphes~3.4.2 for
detector effects. Here we incorporate the default international linear detector card (ILD) of Delphes~3.4.2
for jet reconstruction via anti-$k_T$ algorithm with radius parameter $R = 0.5$ and, for  
the $b$-tagging and misidentification efficiencies of $c$ and light-jets.

\begin{table}[hbt!]
\centering
\begin{tabular}{|c |c| c| c| c | c| c| c |c |c|}
\hline
   \hspace{0.05cm}    BP   \hspace{0.05cm}     & Signal (fb)       & $ZZ$     & $t\bar t$      & $ 4j$    & Total  Bkg. (fb)   \\
\hline
  $a$                                         & 0.304             &  0.18        & 0.112        & 0.461         &   0.753 \\    
                  
\hline
\hline
\end{tabular}
\caption{The signal and background cross sections for BP$a$ in fb for $eeAH$ process at 
$\sqrt{s} = 500$ GeV. 
Total background is presented in the last column.}
\label{bkgcross500}
\end{table}

\begin{table}[hbt!]
\centering
\begin{tabular}{|c |c| c| c| c | c| c| c |c |c|}
\hline    
BP  & Signal (fb)  & $ZZ$ & $t\bar t$ & $ 4j$  & Total Bkg. (fb)\\
 \hline                          
$b$ ($c$)     & 0.199 (0.0015)      &  0.019              & 0.315      & 0.145 & 0.479 \\  
                
\hline
\hline
\end{tabular}
\caption{Same as Table~\ref{bkgcross500} however for $\sqrt{s} = 1$ TeV collision.}
\label{bkgcross1tev}
\end{table}

The events are selected such that it should contain at least four $b$-jets with all having 
$p_T > 20$ GeV and $|\eta| < 2.5$.
The separations between any two $b$-jets should be $ \Delta R>0.4$. 
To reduce the backgrounds further, we demand the scalar sum of $p_T$ of all four $b$-jets
($H_T$) should be $ > 350$ GeV for BP$a$, while for BP$b$ and $c$ we require $> 600$ GeV.
For illustration we show the normalized $H_T$ distributions 
in Appendix for BP$a$ and BP$b$ for $\sqrt{s}=500$ GeV and 1 TeV respectively. 
The signal and backgrounds after selection cuts for 
$\sqrt{s}=500$ GeV and 1 TeV are respectively summarized in Tables~\ref{bkgcross500} and \ref{bkgcross1tev}.

We now estimate the significances from the cross sections summarized in 
Tables~\ref{bkgcross500} and \ref{bkgcross1tev}. 
It is clear that $S/B$ ratios are considerably large for BP$a$ and BP$b$ for the
considered CM energies. Utilizing Eq.~\eqref{poisso} we find that BP$a$ can be discovered at
$\sqrt{s}= 500$ GeV CM energy the with 
$\sim 250$ \fbi integrated luminosity with evidence emerging with as low as $\sim 80$ \fbi data. 
The BP$b$ would require $\sqrt{s}= 1$ TeV run and an evidence may come 
with 120 \fbi but discovery needs 350 \fbi dataset. 
The BP$c$ is below the sensitivity of even $\sqrt{s}= 1$ TeV lepton collider.
Here for all three BPs the signal cross sections
are estimated with $\irbb =0.058$. Therefore we conclude that the nominal value 
for $\rho_{bb}$-EWBG can be fully covered up to $m_A,~m_H\lesssim 200$ (400) GeV with 
moderate integrated luminosity in any future lepton collider if it runs with 
$\sqrt{s}= 500$ GeV (1 TeV) CM energy.

\subsection{The six $b$-jets signature}\label{eeHtohh}
We now discuss a resonant diHiggs production $e^+ e^- \to Z^*\to A H \to A h h$  in future $e^+ e^-$ colliders.
We search this process in which both $h$ decays to $b\bar b$ i.e. in six $b$-jets signature.
Such final state would also receive contribution from process \eehh~which we have considered as well.
For the parameter space described in Table~\ref{bench} only BP$b$ and BP$c$ can facilitate $e^+ e^- \to Z^*\to A H \to A h h$ and \eehh~since  $m_H>2m_h$. 
Note that discovery may already emerge from four $b$-jets signature discussed in previous subsection while
six $b$-jets signature would provide complementarity for $\rho_{bb}$-EWBG.

Based on the LO $Hhh$ coupling given in Eq.~\eqref{lHhh}, we first analyze
the prospect of \eehh~process with both $h$ decays $b \bar b$ i.e., in six $b$-jets
signature with all six $b$-jets having $p_T > 20$ GeV and $|\eta|<2.4$. 
Here we consider two different CM energy $\sqrt{s}=500$ GeV and 1 TeV for illustration.
The CM energies considered would kinematically allow \eehh~process only for BP$b$.
For event generation we follow the same procedure as in \eeZH process i.e. generate events via MadGraph5\_aMC 
followed by hadronization and showering in Pythia 6.4 and adopting default ILD card of Delphes
for fast detector simulation. The corresponding cross sections $\sqrt{s}=500$ GeV (1 TeV) 
before application of any selection cuts reads as $\sim 0.001$ ($\sim0.2$) fb for BP$b$ with $|\irbb|=0.058$.
Following the above mentioned selection cuts, we find 0.0078 fb cross section  
for $\sqrt{s}=1$ TeV, but tiny 0.00003 fb for $\sqrt{s}=500$ GeV. In finding these 
cross sections we have normalized the $\mathcal{B}(h\to b \bar b)$ with the modified
$hbb$ coupling due to nonvanishing $|\irbb|=0.058$.
While no statistically significant cross section is found for 500 GeV run, however one may have $\sim8$ ($\sim24$) events
with 1000 (3000) fb$^{-1}$ integrated luminosity at $\sqrt{s}=1$ TeV. In SM, we find such six $b$-jets backgrounds 
to be negligibly small at $e^+ e^-$ collider, providing ideal environment for discovery of such signature.
This should be compared with the discovery prospect discussed
in Sec.~\ref{eeZHprod} for BP$b$ via \eeZH process, which would 
require  $\sqrt{s}=1$ TeV and $\sim 700$ fb$^{-1}$ data.
In finding the six $b$-jets cross section here we have not included uncertainties
arising from high $b$-jet multiplicity. Hence, we remark that our six $b$-jets cross sections should be treated as exploratory
while a more detailed analysis including possible uncertainties
arising in $e^+ e^-$ collider would be studied elsewhere.

\subsubsection{The vertex correction for $Hhh$ coupling at g2HDM}\label{bbhhvertex}
It is known that one-loop corrections to triple Higgs couplings could be sizable if EWPT
is strongly first order~\cite{Grojean:2004xa,*Kanemura:2004ch,*Senaha:2018xek} 
(for one-loop calculations to the $hhh$ coupling, see also Ref.~\cite{Kanemura:2002vm,*Kanemura:2004mg}). 
Here we clarify if this argument applies for our $Hhh$ coupling. 
Dominant one-loop corrections in the $c_\gamma\to 0$ limit are cast into the form 
\begin{align}
\Delta\lambda_{Hhh} 
&\simeq -\frac{\eta_7}{16\pi^2 v}
\bigg[
3m_H^2\left(1-\frac{\mu_{22}^2}{m_H^2}\right)^2+m_A^2\left(1-\frac{\mu_{22}^2}{m_A^2}\right)^2 \nonumber\\
&\hspace{2cm}+2m_{H^\pm}^2\left(1-\frac{\mu_{22}^2}{m_{H^\pm}^2}\right)^2
\bigg].
\end{align}
Remarkably, the loop correction would not vanish even in the exact alignment 
limit $c_\gamma=0$ due to the presence of the nonzero $\eta_7$, which is in sharp contrast to softly-broken 2HDMs.
In our three BPs, moreover, the loop corrections are constructive since $\eta_7$ is positive
and tree-level $\lambda_{Hhh}$ is negative. In each case of BPs, we find that
\begin{align}
\Delta\lambda_{Hhh}^{\text{BP}a}&\simeq -4.407~\text{GeV},\quad
\Delta\lambda_{Hhh}^{\text{BP}b}\simeq -41.16~\text{GeV},\\
\Delta\lambda_{Hhh}^{\text{BP}c}&\simeq -58.55~\text{GeV}.
\end{align}
One can see that the one-loop corrections are larger than the tree-level values in BP$b$ and BP$c$. 
However, this does not necessarily mean that perturbation breaks down since the tree-level $Hhh$ 
coupling happens to be small by $c_\gamma$, and  moreover, some combinations of quartic couplings 
at one-loop level could be larger than those at tree level though each of quartic couplings is 
less than 4$\pi$ as seen in Table \ref{bench}. As mentioned in Sec.~\ref{const}, the tree-level unitarity is not violated either.
We note that the $H \to f \bar f$ (with $f$ being fermions) decays are not expected to receive large enhancement from the one-loop corrections 
since the two of the three vertices in there are not Higgs-self couplings (for $h\to f\bar{f}$ decays, see, e.g., Ref.~\cite{Kanemura:2019kjg}).
Therefore, $\mathcal{B}(H\to h h)$ would be significantly increased at loop level, leading to much larger possibility 
for discovery at future lepton colliders.

\section{Discussion and Summary}\label{discu}
We have analyzed the prospect of probing EWBG induced by additional bottom Yukawa couplings
at the LHC and future $e^+ e^-$ colliders. We primarily focused on the nominal value $|\irbb| =0.058$ required
for $\rho_{bb}$-EWBG. We show that HL-LHC can offer some probe for such parameter space via \bzh process if 
$300 \lesssim m_A \lesssim 450$ GeV. However, the discovery would be beyond even for HL-LHC. In this regard
we show that future $e^+ e^-$ colliders such as ILC or FCCee would offer exquisite discovery prospect via \eeZH process at
$\sqrt{s}=500$ GeV and 1 TeV. For parameter space where $m_A < m_h +m_Z$, the \bzh process kinematically insensitive but 
a $500$ GeV run of any $e^+ e^-$ collider can discover the $\rho_{bb}$-EWBG via \eeZH process 
with $\sim 250$ \fbi data. The discovery for the same process with heavier $m_A$ is also possible when 1 TeV or larger collision energies are available.

For complementarity, we also studied the prospect \eehh~process in six $b$-jets signature. Based
on our LO order $Hhh$ coupling we found that 1 TeV $e^+ e^-$ collider can indeed discover such a process
as long as $m_H \sim 300$ GeV.
It should be noted that the $Hhh$ coupling could get $\mathcal{O}(100\%)$ one-loop correction 
owing to the sizable Higgs quartic couplings required by the strong first-order EWPT, increasing the significance for the discovery.

We now briefly discuss the impact of turning on other $\rho_{ij}$ couplings. Current direct
and indirect searches still allow $|\rho_{tt}|\sim 0.5$~\cite{Modak:2020uyq} for sub-TeV
$m_A$, $m_{H}$ and $m_{H^\pm}$. Further $\rho_{tc}\sim 0.3$ is also allowed by 
direct and indirect searches and flavor physics~\cite{Hou:2020tnc}. A nonvanishing
$\rho_{tt}$ motivates one to utilize the conventional $gg\to A/H \to t \bar t $~\cite{Aaboud:2017hnm,*Sirunyan:2019wph,*Carena:2016npr} 
and $gg\to t \bar t A/H \to t \bar t t\bar t$~\cite{Kanemura:2015nza,*Craig:2016ygr} 
$gb\to \bar t H^+ \to \bar t t \bar b$ searches~\cite{Plehn:2002vy,Boos:2003yi}. 
For moderate values of $\rho_{tt}$ and $\rho_{bb}$ one may have $gg\to b A /H\to b t \bar t$  signature which 
could be sensitive at the HL-LHC~\cite{Modak:2020uyq}. In this regard it should be reminded that complex $\rho_{tt}$ and 
$\rho_{tc}$ each can account for the observed BAU. Dedicated direct and indirect searches for 
$\rho_{tc}$- and $\rho_{tt}$-EWBG mechanism can be found in 
Refs.~\cite{Kohda:2017fkn,Hou:2018zmg,Hou:2019qqi,Hou:2019gpn,Hou:2019mve,Ghosh:2019exx,Hou:2020tnc,Hou:2018uvr}.
In general if such couplings are real they would not play any role in EWBG, however they would 
aggravate the signatures that we have discussed so far via  suppression in 
the branching ratios of heavy bosons $A/H$. Nevertheless they would open up several additional direct and indirect probes.
Furthermore moderate values of $\rho_{\tau\tau}$  is still allowed by current
data though its impact is not as significant as $\rho_{tt}$ and $\rho_{tc}$. We leave out a detailed discussion of EWBG driven by multiple
$\rho_{ij}$ couplings and subsequent impacts on collider and flavor physics for future work.

As a first estimate, uncertainties arising from factorization scale ($\mu_F$) and
renormalization scale ($\mu_R$) dependences are not included in our LO cross section estimations for \bzh process.
In general, the LO $bg\to bA$ processes have $\sim25-30\%$ scale uncertainties for $m_{A}\sim (300-400)$ GeV 
as discussed in Ref.~\cite{Campbell:2002zm} (see also Refs.~\cite{Dicus:1998hs,Boos:2003yi,Maltoni:2005wd,Harlander:2003ai}). 
In addition it has been found that~\cite{Maltoni:2003pn} the LO cross sections calculated with LO PDF set
CTEQ6L1~\cite{Pumplin:2002vw} have relatively higher factorization scale dependence.
Therefore, we remark that the LO cross sections in our analysis, which we
estimated with LO NN23LO1 PDF set, might have similar uncertainties. A reasonable choice of the factorization scale and  
renormalization scale has been proposed in Ref.~\cite{Maltoni:2003pn}, with $\mu_R=m_A$ and varied from
$\mu_R=m_A/2$ to $\mu_R=2m_A$, along with $\mu_F = m_A/4$ and 
varied from $\mu_F=m_A/8$ to $\mu_F=m_A/2$.
There also exist PDF uncertainties for bottom-quark initiated process as discussed 
in Ref.~\cite{Maltoni:2012pa} (see also Ref.~\cite{Butterworth:2015oua}). 
These would typically  induce some
uncertainties in our results which we leave out for future work.

\vskip0.2cm
\noindent{\bf Acknowledgments.--} \
We thank Kentarou  Mawatari for fruitful discussions and comments.
The work of TM is supported by a Postdoctoral Research Fellowship from 
the Alexander von Humboldt Foundation.\\

\appendix
\section{The normalized $H_T$ distributions for the \eeZH process}
\begin{figure*}[htbp]
\center
\includegraphics[width=.37 \textwidth]{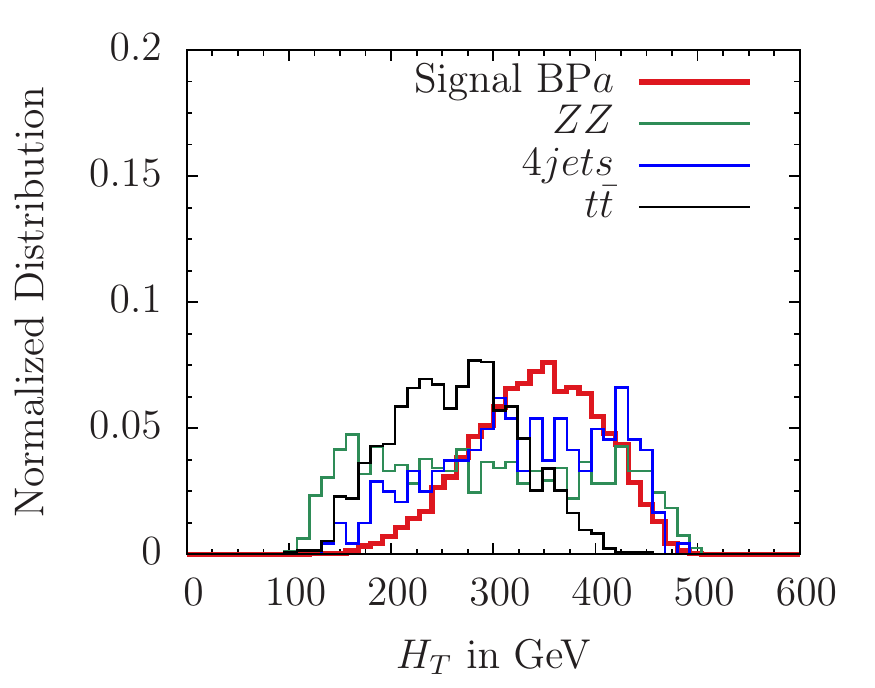}
\includegraphics[width=.37 \textwidth]{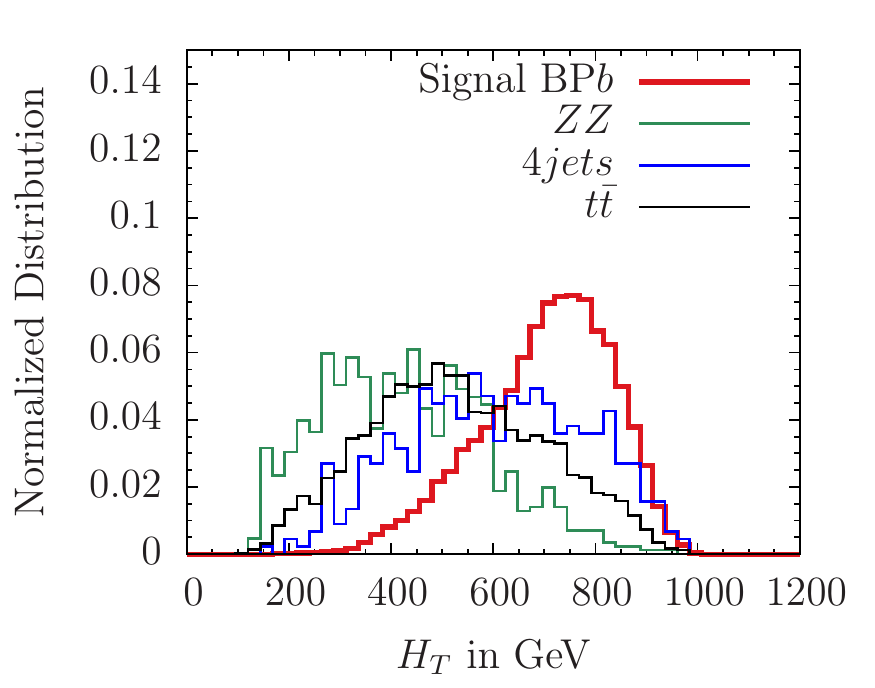}
\caption{
Normalized $H_T$ distributions of the signal and backgrounds for BP$a$ (left) and BP$b$ (right) at $\sqrt{s}= 500$ GeV
and $\sqrt{s}= 1$ TeV respectively.}
\label{HTdist}
\end{figure*}

The normalized $H_T$ distribution is plotted for the \eeZH process in Fig~.\ref{HTdist}.

\section{Theoretical uncertainties of BAU}
It has been known that the so-called vacuum expectation value-insertion approximation (VIA) that we use in our work tends to give overestimated BAU.
We note this issue and add a caveat when interpreting our BAU in Ref.~\cite{Modak:2018csw}. During the review process of this paper, we came across a paper~\cite{Cline:2021dkf} that points out overlooked errors in VIA-based BAU calculations. We have confirmed that our previous estimated BAU has to be divided by 2 after correcting degrees of freedoms of left-handed fermions in calculating BAU. By this correction, we may take $\text{Im}\rho_{bb}\gtrsim 0.058\times2 = 0.116$. We have confirmed that $\text{Im}\rho_{bb}$ can be as large as (0.2-0.25) after taking all the constraints into account (a la Fig.~\ref{paramspace}). However, this is not only the possible revision if we consider theoretical uncertainties described at the last paragraph in this section.

As noted in Ref.~\cite{Cline:2021dkf}, we should be careful about factor 3 when using the strong sphaleron rate in Ref.~\cite{Moore:2010jd} in order to match the correct normalization. In our calculation, however, the strong sphaleron rate is based on Ref.~\cite{Giudice:1993bb} with a corrected color factor and $\kappa=1$, which is consistent in itself. Although the numerical difference between the two estimates happens to be around 3, this discrepancy should be regarded as theoretical uncertainties since $\kappa=1$ is merely a nominal value and systematic error of the lattice calculation could be large~\cite{Moore:2010jd}. 
Another comment on criticism made in Ref.~\cite{Cline:2021dkf} is that $\rho_{bb}$-sourced BAU cannot be estimated by a simple scaling $(\lambda_b/\lambda_t)^2$ from $\rho_{tt}$-sourced BAU, where $\lambda_{t,b}$ are top and bottom SM Yukawa couplings. With the basis-independent CP-violating form~\cite{Guo:2016ixx,Fuyuto:2017ewj}, the scaling factor goes like $(\lambda_b/\lambda_t)(\text{Im}\rho_{bb}/\text{Im}\rho_{tt})$, where the latter factor could be greater than 1.

On this occasion, we recapitulate theoretical uncertainties of our BAU detailed in Ref.~\cite{Modak:2018csw}. In addition to the aforementioned strong sphaleron rate, BAU can be modulated by several factors: (i) bubble wall velocity, (ii) variation of the two Higgs VEVs during EWPT $\Delta\beta$, (iii) critical temperature $T_C$ and corresponding VEV $v_C$, (iv) prescription for UV-divergent piece, (v) CP-conserving source term induced by $\rho_{bb}$, and (vi) an approximation for bubble wall shape. 
The largest uncertainty may come from (ii) since BAU is proportional to $\Delta\beta$. In the minimal supersymmetric SM, $\Delta\beta=\mathcal{O}(10^{-4}-10^{-2})$ depending on $m_A$, while in its extensions $\Delta\beta$ could be as large as $\mathcal{O}(0.1)$. Since detailed study on $\Delta\beta$ is absent in g2HDM at this moment, we take $\Delta\beta=0.015$ for illustration. Our BAU could increase or decrease by an order of magnitude by this factor. The second largest uncertainty could arise from (iv). It is known that temperature-dependent logarithmic divergence exists in the CP-violating source term~\cite{Liu:2011jh,Chiang:2016vgf}. A prescription is to remove it by normal ordering or counterterm~\cite{Liu:2011jh} (for recent criticism on this point, see Ref.~\cite{Kainulainen:2021oqs}). 
Given the fact that this divergence may be attributed to wrong approximation of thermal damping rate, we adopt another prescription in which a momentum integral is made finite using a cutoff (for details, see Ref.~\cite{Chiang:2016vgf}), causing ambiguities in BAU by a couple of factor or more. As for the other uncertainties (i), (iii), (v) and (vi), each of them can increase or decrease BAU by a couple of factor or more.
All in all, theoretical uncertainties are not well under control, and the factor 2 error in our previous BAU could be compensated by, for instance, doubling $\Delta\beta=0.015\times 2$. In conclusion, parameter space for $\rho_{bb}$-EWBG is still open with generous assumptions and awaits more robust BAU calculation.


\bibliography{refs}

\end{document}